\documentclass{aa}
\begin{document}
\thesaurus{03 11.01.2 13.18.1 11.17.3}

\title{High Frequency Peakers: I. The Bright Sample}

\author{D. Dallacasa\inst{1,2} \and C. Stanghellini\inst{3} 
\and M. Centonza\inst{1,2} \and R. Fanti\inst{4,2}}
\institute{ Dipartimento di Astronomia, Via Ranzani 1, I-40127
Bologna, Italy \and Istituto di Radioastronomia - CNR, Via Gobetti 101,  
I-40129 Bologna, Italy. \and Istituto di Radioastronomia - C. P. 169,  
I-96017 Noto (SR), Italy \and Dipartimento di Fisica, Via Irnerio
n. 46, I-40126 Bologna, Italy. }  

\offprints{Daniele Dallacasa, e-mail:  ddallaca@ira.bo.cnr.it}

\date{Received 11 July 2000; accepted 9 October 2000}

\maketitle
\markboth{High Frequency Peakers: I. The Bright Sample}{Dallacasa et al.}

\begin{abstract}

Here we present a sample of sources with convex radio spectra peaking
at frequencies above a few GHz. We call these radio sources High
Frequency Peakers (HFPs). This sample extends to higher turnover
frequencies than the samples of Compact Steep Spectrum (CSS) and GHz
Peaked Spectrum (GPS) radio sources. HFPs are rare due to the
strong bias against them caused by their turnover occurring at
frequencies about one order of magnitude higher than in CSS-GPS
samples.

The sample has been selected by a comparison between the Green Bank
survey (87GB) at 4.9 GHz and the NRAO VLA Sky Survey (NVSS) at 1.4
GHz. Then the candidates have been   
observed with the VLA at 
1.365, 1.665, 4.535, 4.985, 8.085, 8.485, 14.96 and 22.46 GHz in order
to derive a simultaneous radio spectrum, and remove variable sources
from the sample. The final list of genuine HFP sources consists of 55
objects with flux density exceeding 300 mJy at 4.9 GHz at the time of
the 87GB observation. Optical identifications are available for 29 of
them; 24 are high redshift quasars, 3 are galaxies (one of them has
indeed broad lines in the optical spectrum) and 2 are
BL Lac objects. The remaining sources are mostly empty fields (17) on 
the digitised POSS or have uncertain classification (9).

\end{abstract}

\section{Introduction}

GHz-Peaked Spectrum and Compact Steep Spectrum radio sources are
identified with both galaxies and quasars; however the latter are
likely to represent a different phenomenon (Stanghellini et
al. \cite{Stang96}; Snellen et al. \cite{Snellen99}), and at least
partially affected by Doppler boosting (Fanti et al. \cite{Fanti90})
as can be inferred   also by their pc-scale radio morphologies,
generally more similar to those found in flat spectrum, variable
sources. However a few quasars have radio structures similar to those
found for CSS/GPS galaxies (Dallacasa et al. \cite{Dallacasa95},
Stanghellini et al. \cite{Stang97b}).

The pc-scale radio morphology derived from VLBI observations can be
used to evaluate the effects of Doppler boosting, that is likely to play
a role in sources dominated by a single, unresolved component, in
which a combination of opacity effects and speed of the plasma flow
might produce peaked radio spectra (see Snellen \cite{SnellenPhD},
Snellen et al. \cite{Snellen99}). Sources with their radio axis close
to the plane of the sky  instead are characterised by a weak core
accounting for a small fraction of the total flux density.

In the framework of individual, powerful radio $galaxy$ growth,
GPS and then CSS radio sources are nowadays considered the early
stages, as the radio emitting region grows and expands within the
interstellar matter of host galaxy, before plunging into the
intergalactic medium to originate the extended radio source population
(Fanti et al. \cite{Fanti95}; Readhead et al.\cite{Readhead96}; Begelman
\cite{Begelman96}; Snellen et al. \cite{Snellen00}). Their radio
structure is dominated by jets, lobes and hot-spots, while the cores
generally account for a relatively small fraction of the total flux
density. The morphologies are reminiscent of the FR~II radio galaxy
class, and they also share the same radio power range.  

It has been shown (Owsianik \& Conway \cite{OC98}; Owsianik at
al. \cite{OCP98}) 
that in the small double lobed Compact Symmetric Objects (CSOs) the
projected separation speed of the outer edges (hot-spots) is about
0.2c, confirming this hypothesis. The dynamical age of sources of
about 50-100 pc in size is of the order of $10^3$ yr. The estimates of
the radiative ages of the small radio sources are consistent with the
hypothesis that they are young (Murgia et al. \cite{Murgia99}). 

There is a correlation between (projected) linear size and turnover
frequency (O'Dea \cite{Odea98}). As the radio source expands the turnover
moves to lower frequencies as the result of a decreased energy density 
within the radio emitting region. The turnover in the radio
spectrum is either due to synchrotron self-absorption within the small
radio emitting regions or to free-free absorption in the ionised
region surrounding the radio source. In some cases it is also possible
that a combination of these two effects is required.

The samples of powerful CSS and GPS radio sources studied so far
(e.g. Fanti et al. \cite{Fanti90}; Stanghellini et al. \cite{Stang98})
list the brightest sources with turnover frequencies ranging from
about 100 MHz to about 5 GHz. The same distribution in peak frequency
can be found in the samples of somewhat less powerful objects by
Snellen et al. \cite{Snellen98} and Marecki et al. \cite{Marecki99}).

A sample of objects with turnover frequencies above 5 GHz would
represent $smaller$ and therefore $younger$ radio sources. We call
these sources ``High Frequency Peakers'' (HFPs). They are rarely
found in CSS and GPS samples since their spectral turnover occurring
between a few and a few tens of GHz, generally makes them
relatively weak at the frequencies where radio catalogues are available.

This class of sources also plays an important role when considering
the contribution of discrete sources to the Cosmic Microvawe
Backgorund (CMB) (De Zotti et al. 2000). The density and the power of
the HFP sources have to be taken into serious account in space missions
like MAP and PLANCK aiming to produce high resolution and high
sensitivity ($\mu\,K$) images of the CMB radiation.

\section{Candidate High Frequency Peakers}
\label{structure}
CSS and GPS sources have convex radio spectra peaking at frequencies
ranging from about hundred MHz to a few GHz; we assume/define that
HFPs have similar spectral properties, with the peak occurring at a
few GHz or higher frequencies. 

The availability of large areas covered by radio surveys, and the need
of a spectral peak at high frequencies made the choice of the NVSS
(Condon et al. \cite{Condon98}) and of the 87GB (Gregory et
al. \cite{Gregory96}) catalogues quite natural.  

We cross correlated the 87GB catalogue at 4.9 GHz with the NVSS
catalogue at 1.4 GHz and selected the sources with inverted spectra,
and in particular those with a slope steeper than $-$0.5
($S\propto\nu^{-\alpha}$). 
We defined {\bf two} samples of candidates: the ``{\bf bright}''
sample, with sources brighter than 300 mJy at 5 GHz and covering
nearly the whole area of the 87GB (declination between 0 and
+75$^\circ$), excluding objects with $\mid{b_{II}\mid}<10^\circ$ to
avoid the galactic plane and ease the optical identification work; the
``{\bf faint}'' sample is restricted to the area covered by the FIRST
survey (Becker et al. \cite{Becker95}) as well, and limited to sources
brighter than 50 mJy at 4.9 GHz. This work presents the ``bright''
sample, while a forthcoming paper (Stanghellini et al. \cite{StangP})
will describe the ``faint'' sample.   

The search for candidates
for the bright sample started with the 1795 sources from the 87GB
stronger than 300 mJy and in the region of the sky described above. 
We used a simple fortran program to make a cross correlation of the
positions of these sources with the catalogue derived from the
NVSS. The error in the position for the sources in the 87GB is 
much larger than that associated with the NVSS, and is generally of
the order of 10-15 arcseconds in both RA and DEC. We considered
positionally coincident the sources with difference in either
coordinate ($\Delta_{RA}$ or $\Delta_{DEC}$) smaller than the largest
between 45 arcsec and 3 times the error reported in the 87GB
catalogue. Only 55 sources could not be identified (3.1\%) since they
fall in areas not yet covered by our release of the NVSS catalogue (as
of July, 1999). Among the remaining 1740 objects, 164 (9.4\%) had an
inverted spectra with slope steeper than $-0.5$ and they were included
in our starting list. 
We then inspected the NVSS images to make sure that the component in
the catalogue accounted for the whole flux density. The extended
objects (typically FRII and a few FRI or complex radio sources),
resolved by the NVSS but a single source in the 87GB, were removed. 
Our list of ``bright'' HFP candidates consisted of 103 sources. After
a search for optical identification one source have been dropped,
since it is associated with a planetary nebula (\object{J1812+0651}),
hence to a 
completely different physical phenomenon. Therefore, the final list is
made up with 102 (5.9\% of the starting dataset) candidates and is
presented in Table 1: column 1  gives the J2000 name; columns 2 and 3
provide the J2000 coordinates from the NVSS catalogue (very accurate
positions can be found in the Jodrell Bank VLA Astrometric Survey,
JVAS catalogue; Patnaik et al. \cite{Patnaik92}, Browne et
al. \cite{Browne98}, Wilkinson et al. \cite{Wilkinson98}); columns 4, 5
and 6 report the flux densities in the NVSS, 87GB and JVAS catalogues,
respectively; column 7 shows the spectral index between the NVSS and
the 87GB; column 8 reports whether the source belongs to other
relevant samples (see below); columns 9, 10 and 11 give the optical
ID, magnitude and redshift; finally column 12 provide the B1950 source
name.   

We found 14 sources in common with the new GPS candidate starting list
in Marecki et al. (\cite{Marecki99}) ('m' in column 8), three objects
are in the 'bright' GPS sample from Stanghellini et
al. (\cite{Stang98}) ('st'), two sources in the 'faint' GPS sample
from Snellen et al. (\cite{Snellen98}) ('sn'). It has been useful to
compare our list to the 550 compact extragalactic object in Kovalev et
al. (\cite{Kovalev99}) ('K'), where nearly simultaneous radio spectra
are available between 1 and 22 GHz.  We searched our HFP candidates in
the Caltech-Jodrell Bank flat-spectrum sample (Taylor et
al. \cite{Taylor96} and references therein) ('pr', 'cj1' and 'cj2')
and with the  Kellermann et al. (\cite{Kellermann98}) ('k') 
VLBA survey at 15 GHz, in order to have images of the pc-scale radio
morphology. Finally most of our HFP candidates have very short
snapshot images in the VLBA Calibrator Survey (VCS, Peck \& Beasley
\cite{Peck98}). 

The optical ID (capitals) and redshift are from the NED database, when
available. We also report our optical ID on the digitised red plates
of the Palomar Observatory Sky Survey (POSS) (small 'g' and 's' for
extended or stellar) when no other optical information is available; a
'?' following the optical identification means that the classification
is uncertain. We remark that the optical magnitudes, mostly from the
NED database, reported in Table 1 are not homogeneous (i.e. in the
same band). Also variability plays an important role, given that a
significant fraction of the sources are associated with blazars. In
fact among the candidates there are also 3 BL Lac objects from the 1
Jy sample (Stickel et al. \cite{Stickel91}) and a few other sources
known to be variable. 

\section{The simultaneous radio spectra}

Simultaneous multifrequency observations are necessary to remove flat
spectrum variable sources from the sample; in fact our selection
criteria are based on two observations at 4.9 and 1.4 GHz taken 
a few years apart. Variable sources that happened to be in a ``high''
activity state at the time of the 4.9 GHz observation are selected by
our criteria, and need to be removed.

Hence, we have observed at the VLA the whole ``bright'' sample. 
We carried out nearly simultaneous flux density measurements at L band
(with the two IFs at 1.365 and 1.665 GHz), C band (4.535 and 4.985
GHz), X band (8.085 and  8.485 GHz), U band (14.935 and 14.985 GHz)
and K band (22.435 and 22.485 GHz). The observing bandwidth was chosen
to be 50 MHz per IF except at 1.665 GHz, set to 25 MHz in order to
avoid radio interference. 
  
Each source was observed typically for 40 seconds at each frequency in
a single snapshot, cycling through frequencies. This means that our
flux density measurements are nearly simultaneous, given that they are
separated by 1 minute apiece. 

For each observing run we spent one or two scans on the two primary
flux density calibrators \object{3C286} or \object{3C147}. Secondary
calibrators were 
observed for 1 minute at each frequency about every 25 minutes; 
they were chosen aiming to minimise the telescope slewing time and
therefore we could not derive accurate positions for the radio sources
we observed. Accurate positions can be obtained from the JVAS
catalogue (Patnaik et al. \cite{Patnaik92}, Browne et
al. \cite{Browne98}, Wilkinson et al. \cite{Wilkinson98}). 

Information on the date and duration of the observing runs is summarised
in Table 2. Sources from the bright and faint samples were observed
together in each run in order to optimise the observing schedule.  

The data reduction has been carried out following the standard procedures
for the VLA implemented in the NRAO AIPS software. Separate images
for each IF were obtained at L, C and X bands in order to improve the
spectral coverage of our data.
Imaging has been quite complicated at L band since a number of
confusing sources fall within the primary beam, and an accurate flux
density measurement could be obtained only once the confusing sources
had been cleaned out. Generally one iteration of phase-only 
self-calibration have been performed before the final imaging. On the
final image we perfomed a Gaussian fit by means of the task JMFIT, and
also measured the source flux density with TVSTAT and
IMSTAT. Generally all the HFP candidates were unresolved by the
present observations; when JMFIT found some extension, it was
generally much smaller than the beam size, and therefore we did not
consider it. 

The r.m.s. noise levels in the image plane is relevant only
for measured flux densities of a few mJy, which is not the case for
our sources; in fact the major contribution comes from the amplitude
calibration error. 

At K band, the antenna gain and atmospheric opacity vary significantly
with elevation, introducing visible effects below $40^\circ$,
progressively reducing the correlated flux density. Only two observing
runs (19Dec98 and 14Jun99) had a few sources observed at low
elevations. The flux densities were then readjusted by comparing the
flux density of \object{J0111+3906} (\object{B0108+388}), a well known
supposedly non variable GPS source, observed in a run close in time. 

At the end, we estimate that the overall amplitude error (1 sigma) is
3\% at L,C and X bands, 5\% at U band and finally 10\% at K band.

\section{The ``Bright'' HFP Sample}

We derived the spectral indices between any pair of adjacent
frequencies, and then we classified the sources into two sets:
1) the genuine HFPs, i.e. sources with peaked radio spectrum and at
least a spectral index below -0.5 (55 objects) 
2) flat spectrum sources (48). These objects with their flux densities
are listed in Table 3 and Table 4 respectively, according to their
classification. 

It is interesting to note that also among these genuine HFP sources,
flux density variability is not uncommon (see Section 7.1), and indeed
12 objects have flux densities below 300 mJy at 4.9 GHz, and at the
time of our observations would have been dropped from the bright sample.

Our selection picked up also three well known GPS sources, namely
\object{J0111+3906} (G), \object{J1407+2827} (G) and
\object{J2136+0036} (Q) (Stanghellini et al. \cite{Stang98}). In fact
their radio spectra peak between 4 and 10 GHz. 

Of the two GPS sources in Snellen et al. (\cite{Snellen98}),
\object{J1623+6624} turned out to be slightly below the 300 mJy limit,
(Table 3 and Figure 1), while \object{J1551+5806} in our simultaneous
data has a flat spectrum (Table 4 and Figure 2) and therefore has been
classified as a non-HFP source. 

\object{J1751+0939} is a well known BL Lac object from the 1Jy sample
(Stickel et al. \cite{Stickel91}), and it is classified as a HFP
source. It was observed twice, and at both epochs the radio spectrum
meets our selection criteria. Another BL Lac object
(\object{J0625+4440}) has a genuine HFP radio spectrum. 
This confirms that it is possible that beamed radio sources
like BL Lac objects, or more generally blazars, posses radio spectra
peaking above a few GHz, as the result of a self-absorbed synchrotron
emission from the jet base. The sample of HFP sources presented here
is therefore expected to collect a mixture of Doppler boosted objects
together with the young pregenitors of the GPS-CSS population. The
determination of the pc-scale morphology will be an important tool to
distinguish among these two classes.

It is interesting to analyse the the 23 sources in common with the
Kovalev et al. (\cite{Kovalev99}) list of compact extragalactic
objects, given that they provide nearly simultaneous flux density
measurements between 1 and 22 GHz, and our selection criteria can be
applied in the same way. Among the 16 sources we define genuine HFPs
only 11 are HFPs in Kovalev et al.,
while 5 would have been classified as flat spectrum sources. On the
other hand, among the remaining 7 non-HFPs, 4 would be HFPs
(\object{J0424+0036}, \object{J0811+0146}, \object{J1146+3958} and
\object{J2321+3204}) according to Kovalev et al. measurements.  This
could be due to a progressive fall off of the radio spectrum at low
frequencies, since the lowest frequency in Kovalev et al. is about
40\% lower than our, or to an effective variability in this class of
sources, both in flux density and in spectral shape (see also
Sect. 7). We would like to remark that the BL Lac object
\object{J1751+0939} is a HFP source also in Kovalev et al. data. 

There is no clear segregation on the fraction of HFPs based on the
optical identification. In fact the number of HFPs versus total is 4/9
for galaxies, 23/36 for quasars, 2/8 for BL Lacs, 16/32 for empty
fields and 10/25 for objects with uncertain classification.
The fractions of empty fields and identifications with an uncertain
classification are still too large to allow a proper statistical
analysis. An 
optical identification program is in progress at the 3.6m Telescopio
Nazionale Galileo (TNG) in La Palma.

\section{Spectral peaks}

We fitted the simultaneous radio spectra of the genuine HFP sources in
Table 3 in order to estimate the spectral peak and the turnover
frequency. We first used the function reported as Eq. (1) in Snellen
et al. (\cite{Snellen98}) or Eq. (2) in Marecki et
al. (\cite{Marecki99}) in which some assumptions about the physics in
the radio sources are taken into account to derive the spectral
shape. However, Snellen et al. (\cite{Snellen98}) 
used the function only for determining the peak frequency, flux
density peak and the Full Width Half Maximum of their fitted spectra. 
Indeed, it is well known that the description of the radio spectrum in
terms of a single homogeneous synchrotron component is too simplistic,
and often the parameters derived from spectral peaks can be taken as
gross estimates only.

Our simultaneous radio spectra have a much better sampling and
more uniform uncertainties than in Snellen et al. (\cite{Snellen98})
and Marecki et al. (\cite{Marecki99}). We tried a fit with the
forementioned function, but we also  fitted the radio data
with a purely analytic function, with no physics behind, given it is
used to determine only ``analytical'' quantities, namely the peak and
the frequency at which it occurs. Rearranging the parameters from
Kovalev et al. (\cite{Kovalev99}) we used the following function: 

$$Log~S~=~a~-~\sqrt{b^2+(cLog(\nu)-d)^2}$$

From this fitting curve we derived the spectral peak ($S_m$ and
$\nu_m$), representing the actual maximum, regardless the point where
the optical depth is unity, or any other physical measure. The
parameters $a, b, c$ and $d$ are purely numeric, and do not provide
any direct physical information. 

For two sources (\object{J0519+0848}, first epoch, and
\object{J2257+0243}) we excluded the two flux densities at 1.365 and
1.665 GHz in order to have a good fit of the spectral peak. 

It is interesting to note note that the peak frequencies in our bright
HFP sample are about a factor of 5 higher than in Snellen et
al. (\cite{Snellen98}), and our spectra generally appear broader. We
do not provide the FWHM of our fitted spectra since the range sampled
by our simultaneous measurements is rather small, and very seldom
the 22 GHz flux density falls below half the peak flux density, making
the detemination of the width rather uncertain. We did not consider
datapoints at low (WENSS, Rengelink et al. \cite{Rengelink97}; Texas,
Douglas et al. \cite{Douglas96}) and high frequencies (e.g. Steppe et
al. \cite{Steppe88}, \cite{Steppe92}, \cite{Steppe93},
\cite{Steppe95}) since one of the key point in the selection is the
simultaneous measurements at the various frequencies. 

We remark further that only a few sources (namely \object{J0217+0144},
\object{J0357+2319}, \object{J1645+6630}, \object{J2101+0341} and
\object{J2123+0535}) have flux densities still rising at 22 GHz,
although all of them have slopes with a spectral index close to
0. Also the first epoch of \object{J1016+0513} has a 
spectral peak above 22 GHz; this source indeed possess a rather
unusual radio spectrum, and also show prominent variability in both
flux density and spectral shape. In fact the peak flux does not change
very much but the peak frequency moves from about 22 GHz down to 7.1
GHz in the second epoch; the flux density at 5.0 GHz, close to the
87GB, rises from 0.20 up to 0.51 Jy.

A spectral shape similar to the first epoch for \object{J1016+0513}
has been found in \object{J2257+0243}, whose spectral peak is 0.58 Jy
at about 20 GHz, but its flux density at 4.9 
GHz is only 0.27 Jy, and therefore below the 87GB limit. 

GPS sources have steep optically thin radio spectra, and they are not
as broad as found in the majority of our genuine HFPs. We 
would need observations at higher frequencies in order to study the
optically thin emission, and to disentangle the contamination by
beamed objects, since we do expect that beamed objects are
characterised by a dominant flat spectrum component (on the pc scale),
whose relative relevance increases with frequency.

\section{Comparison with NVSS, 87GB and JVAS flux densities}

We compared the flux densities we measured with the values from 
the NVSS at 1.4 GHz, from the 87GB at 4.9 GHz and also from the JVAS
catalogue at 8.4 GHz, where all our candidates except
\object{J0037+0808} and \object{J1424+2256} are included. We always
considered our data at the closest frequency to that of the
forementioned catalogues. 

The effects due to confusion are generally marginal since we can
evaluate any contribution from confusing sources in the large beam of
the 87GB at 4.9 GHz from the NVSS image and also from the VLA image
obtained from our data. 

Only one case (\object{J0116+2422}) needs attention since there are
two unresolved NVSS sources separated by about 73 arcsec in p.a. $\sim
60^\circ$, but within the 87GB beam. By summing the two flux densities
at 1.4 GHz, the spectral index with the 87GB would have dropped the
source out of the candidate sample ($\alpha=$0.43). The 87GB position
lies in between the two NVSS sources, at 35 arcsec from the
southwestern one. However we decided to observe anyways this pair,
since if one would have had a commonly steep spectrum, the other would
have been inverted enough to be an HFP candidate. 

Figure 3 displays the comparisons between our flux density measurements
and those from the foremetioned catalogues. We distinguished among HFP
and non-HFP sources by using filled or empty squares
respectively.

At 1.4 GHz the data are evenly scattered around the 1:1 relation
without any evidence for a different behaviour between the two classes
of HFP and non-HFP sources. If we consider $R$ as the ratio between
our flux density over the measure in the catalogue (NVSS, 87GB or
JVAS) at about the same frequency, we find that the median of the
$R_{1.4}$ distribution is 0.99 for the 55 HFPs and 0.94 for the 47
non-HFPs. 

At 4.9 GHz it is clear that, on average, the 87GB flux densities
exceed our measurements; it is also evident that genuine HFP and
non-HFP objects behave completely different, with the former nearly
randomly scattered around the 1:1 relation (the median of the $R_{4.9}$
distribution is 0.97), and the latter generally well below (the median
is 0.51). As mentioned in Section 3, this has to be expected since
our selection criteria favours the inclusion of variable sources at a
high state at the time of the 87 GB observation, and therefore it is
natural to find non-HFP sources in the lower right part of the panel. 

At 8.4 GHz the scenario is completely different, with the non-HFP
sources generally weaker in our observations than in the JVAS
catalogue (the median of $R_{8.4}$is 0.71), but with the HFP sources
$brighter$ (median~=~1.16). 

We should consider that the comparisons of the flux densities at 4.9
GHz might be somehow influenced by a different technique for the
observation and by a totally different instrument, but it should not
affect the class of (point) source observed. Therefore the
significantly different behavior between HFPs and non-HFPs is
physically relevant.

A proper statistical analysis will be carried out in a forthcoming
paper where also the 'faint' sample will be considered (Stanghellini
et al. 2001).

\section{Discussion}

Our constraints in declination and galactic latitude leave an area of
4.978 sr; however the effective area where we have searched for cross
identification between the 87GB and the NVSS is 4.825 sr, if we
consider that 3.1\% of the sources fall in regions where the NVSS is
not available as yet. We ended up with 55 genuine ``bright'' HFP
sources and this gives a density of 11.4 sources per sr. 
We will discuss in more detail the source counts as a function of the
limiting flux density after the selection of the ``faint'' HFP sample
in order to span a wider flux density range, with a larger number of
objects. In the framework of the ``youth'' model for CSS and GPS radio
sources the number of CSSs/GPSs/HFPs is related to the time
they spend in each stage and to the variation of the total radio
luminosity with time. It is therefore very likely that the number of
HFP progenitors of GPS sources is small, i.e. only a fraction of the
``bright'' HFP sources will indeed develop into GPS and CSS stages.
A further multifrequency radio observation, the determination of the
VLBI structure, the optical identification and spectroscopy will be
important tools to disentangle the intrisically young radio sources
from the beamed objects.

\subsection{Size estimate}

We have derived estimates for the component size by assuming that the
radio emission is originated in a single homogenous region, and that
the source is in equipartition. Following Scott \& Readhead
(\cite{Scott77}) and assuming an injection index of 0.75, we derive
component sizes in 
the range from 0.3 to 3 mas. In a similar way we have estimated the
source largest size (LS) from the relation between the source size
(considering the outer edges) and the turnover frequency derived by
O'Dea (\cite{Odea98}) for bright CSS and GPS sources. The values
obtained cover a range between a few mas to a few tens of
mas. Therefore, these objects would represent the extreme
representatives of the GPS population, being on average about one
order of magnitude smaller. 
Hence, in the framework of a radio source growing and expanding within
the host galaxy, HFPs would represent the very early stage.

\subsection{Flux density variability}

Figure 3, can also be used to test for flux density variability: it is
rather evident that this phenomenon is common among HFP sources. This
is also reinforced by the simultaneous radio spectra shown in Figure 1
where often the NVSS, 87GB and JVAS datapoints clearly stands
out with respect to our measurements.

Further, a few sources have been observed twice or more in our runs,
and the case of the forementioned \object{J1016+0513} is an example of
significant variability in both spectral shape and flux density.
If we assume that the spectral peak has moved from about 22 GHz down
to 7.1 GHz as effect of pure expansion, the source component
equipartition size would have increased by a factor of 2.9 from
December 1998 to October 1999; assuming a redshift of 1.0 the size
would have grown from 0.24 to 0.71 mas corresponding to 1.0 and 3.0 pc
(with $H_o=100~$km~s$^{-1}$Mpc$^{-1}$ and $q_o=0.5$), and leading to an
expansion velocity of 15.6$c$! This suggests/confirms either
that a significant contamination of the HFP sample by beamed objects
or that the sources are not in equipartition.

Other sources like \object{J0927+3902}, \object{J1751+0939} and
\object{J2136+0041} only show flux density variability, without an
appreciable peak frequency or shape change. Finally  other sources
like \object{J0111+3906}, \object{J0428+3259}, \object{J1045+0624}, 
\object{J1407+2827} have turned out to have constant flux densities
within the calibration errors. In particular \object{J1407+2827}
(alias \object{OQ208} or \object{B1404+286}) has an interesting
history of decrease of its flux density by about 20\% at cm
wavelenghts in the eighties (Stanghellini et al. \cite{Stang97a}), but
remained stable since then.  

All these results deserve a deeper investigation by means of
a second VLA multifrequency observation.

\section{Summary}

We have presented a sample of 55 High Frequency Peakers,
i.e. radio sources with radio spectra having their maxima at
frequencies about an order of magnitude higher than known GPS
samples. This sample is intended to provide smaller and younger
radio sources, but it is also likely to contain objects with properties
different from the conventional CSS-GPS class. Further work is
required to complete the optical identification and redshift
determination, while further multifrequency polarimetric VLA
observations would provide useful insights to distinguish subclassed
within our HFP sample. The determination of the pc-scale
morphology will be a key point to distinguish between ``young'' and
beamed objects.

Another HFP sample on a restricted area but deeper by a factor of 6 in
limiting flux density will be presented in a forthcoming paper
(Stanghellini et al. 2001). 

\begin{acknowledgements}
We acknowledge financial support from the Italian M.U.R.S.T., under the
program Cofin-02-98. The National Radio Astronomy Observatory is
operated by Associated Universities, Inc. under cooperative agreement
with the National Science Foundation. We acknowledge the use of the
NASA/IPAC Extragalactic Database (NED), which is operated by the Jet
Propulsion Laboratory (JPL), California Institute of Technology, under
contract with the Natiional Aeronautics and Space Administration
(NASA). Finally we wish to thank the referee, Ignas Snellen, whose
comments considerably improved the manuscript.
\end{acknowledgements}

{\small

\begin{table*}
\caption{Candidates observed with the VLA. All columns are
self-explicative, except column 8, where a reference to other samples
is reported; a full description is given in the text. }
\begin{center}
\begin{tabular}{cccrrccccclc}
\hline 
\hline 
J2000 &R.A.(J2000)&Dec.(J2000)&$_{NVSS}$&$_{87GB}$&$_{JVAS}$&&Other&ID&&~~~z&B1950\\ 
Name  &h~m~s  &d~m~s   & mJy  & mJy & mJy &$\alpha_{1.4}^{4.9}$&sampl.&&&&  Name  \\
\hline 				
0003+2129&00~03~19.34&21~29~44.5&  83.7&  352& 259&-1.17 &       &   &    &       &0000+212\\
0005+0524&00~05~20.21&05~24~10.1& 127.1&  300& 235&-0.70 &       & Q &16.2& 1.887 &0002+051\\
0037+1109&00~37~26.02&11~09~50.4& 235.4&  456& 232&-0.54 &       &   &    &       &0034+108\\
0037+0808&00~37~32.15&08~08~12.6&  96.9&  320&    &-0.97 &       &   &    &       &0034+078\\
0039+1411&00~39~39.63&14~11~58.0& 266.3&  504& 386&-0.52 &       &   &    &       &0037+139\\
0107+2611&01~07~47.88&26~11~10.1& 189.7&  364& 323&-0.53 &       &   &    &       &0105+259\\
0111+3906&01~11~37.31&39~06~27.6& 429.0& 1321& 830&-0.92 &pr,st,K& G &22.0& 0.668 &0108+388\\
0116+2422&01~16~33.47&24~22~14.8& 154.8&  457& 226&-0.88 &       &   &    &       &0113+241\\
0132+4325&01~32~44.00&43~25~32.0& 150.9&  347& 235&-0.68 &cj2,m  & S &18.7&       &0129+431\\
0205+1444&02~05~13.11&14~44~32.2& 197.8&  365& 180&-0.50 &       &   &    &       &0202+145\\
0217+0144&02~17~48.93&01~44~48.9& 750.8& 1608&1176&-0.62 &       & Q &18.3& 1.715 &0215+015\\
0254+3931&02~54~42.77&39~31~33.5& 199.3&  408& 367&-0.58 &cj2,m  & G &17.0& 0.289 &0251+393\\
0310+3814&03~10~49.95&38~14~53.5& 236.9&  760& 453&-0.95 &cj2,m  & Q &18.5& 0.816 &0307+380\\
0312+0133&03~12~43.56&01~33~17.1& 459.6& 1033& 474&-0.66 &       & Q &18.2& 0.664 &0310+013\\
0313+0228&03~13~13.36&02~28~34.9& 174.8&  322& 127&-0.50 &       & g?&20.1&       &0310+022\\
0329+3510&03~29~15.35&35~10~08.1& 262.4&  545& 408&-0.60 &       &   &    &       &0326+349\\
0357+2319&03~57~21.63&23~19~53.5& 176.6&  327& 340&-0.50 &       &   &    &       &0354+231\\
0424+0036&04~24~46.84&00~36~06.6& 493.5& 1118& 288&-0.67 &  K    &BL &17.0& 0.310 &0422+004\\
0428+3259&04~28~05.82&32~59~52.0& 152.3&  589& 509&-1.10 &       &   &    &       &0424+328\\
0509+0541&05~09~25.96&05~41~35.7& 536.4& 1026& 684&-0.53 &       & S &15.3&       &0506+056\\
0519+0848&05~19~10.77&08~48~57.0& 202.8&  420& 160&-0.59 &       &   &    &       &0516+087\\
0530+1331&05~30~56.43&13~31~55.2&1556.7& 2995&3110&-0.53 &  k,K  & Q &20.0& 2.060 &0528+134\\
0559+5804&05~59~13.39&58~04~03.9& 392.3&  906& 501&-0.68 & cj2   & Q &18.0& 0.904 &0554+580\\
0625+4440&06~25~18.27&44~40~01.7& 122.7&  369& 183&-0.90 & m     &BL &    &       &0621+446\\
0638+5933&06~38~02.85&59~33~22.2& 254.2&  482& 553&-0.52 & cj2   &   &    &       &0633+595\\
0642+6758&06~42~04.23&67~58~35.5& 192.9&  499& 436&-0.77 & cj2   & Q &16.5& 3.180 &0636+680\\
0646+4451&06~46~32.09&44~51~16.8& 452.8& 1191&2184&-0.79 &cj1,k  & Q &18.5& 3.396 &0642+449\\
0650+6001&06~50~31.21&60~01~44.5& 472.7&  920& 753&-0.54 & cj1   & Q &18.9& 0.455 &0646+600\\
0655+4100&06~55~10.03&41~00~10.6& 226.1&  425& 373&-0.51 &cj2,m  & G &14.6&0.02156&0651+410\\
0722+3722&07~22~01.26&37~22~28.6& 150.2&  306& 234&-0.58 & m     & s &17.5&       &0718+374\\
0733+0456&07~33~57.45&04~56~14.1& 219.3&  555& 232&-0.76 &       & g?&19.5&       &0731+050\\
0811+0146&08~11~26.68&01~46~54.5& 535.8& 1469&1310&-0.82 & k,K   &BL &17.5&       &0808+019\\
0831+0429&08~31~48.88&04~29~38.5&1155.9& 2136&1235&-0.50 & k,K   &BL &16.5& 0.180 &0829+046\\
0854+0720&08~54~35.08&07~20~24.2& 136.7&  309& 166&-0.66 &       &   &    &       &0851+075\\
0927+3902&09~27~03.03&39~02~20.7&2885.1& 7480&8012&-0.78 &k,pr,K & Q &17.9& 0.6948&0923+392\\
0958+6533&09~58~47.22&65~33~54.2& 729.9& 1417&1206&-0.54 &       &BL &16.7& 0.368 &0954+658\\
1016+0513&10~16~03.11&05~13~03.5& 401.7&  745& 303&-0.50 &       & S &20.0&       &1013+054\\
1018+0530&10~18~27.82&05~30~29.8& 278.3&  652& 295&-0.69 &       & S &20.5&       &1015+057\\
1033+0711&10~33~34.00&07~11~26.2& 155.9&  364& 216&-0.69 &       & Q &... & 1.535 &1030+074\\
1045+0624&10~45~52.74&06~24~36.2& 157.3&  457& 397&-0.87 &       & Q &17.9& 1.507 &1043+066\\
1048+7143&10~48~27.56&71~43~35.2& 736.8& 2410&1259&-0.96 & cj1   & Q &19.0& 1.150 &1044+719\\
1056+7011&10~56~53.70&70~11~46.0& 310.9&  675& 603&-0.63 & cj1   & Q &18.5& 2.492 &1053+704\\
1146+3958&11~46~58.31&39~58~34.9& 331.4&  739& 565&-0.65 & cj1,K & Q &18.0& 1.088 &1144+402\\
1148+5254&11~48~56.63&52~54~25.7&  93.4&  304& 597&-0.96 & cj2   & Q &15.5& 1.632 &1146+531\\
1209+4119&12~09~22.81&41~19~41.0& 274.2&  515& 486&-0.51 &cj2,m  & S &16.3&       &1206+416\\
1228+3706&12~28~47.40&37~06~12.0& 383.8&  953& 868&-0.74 & cj2   & G &18.0& 1.515 &1226+373\\
1302+5748&13~02~52.47&57~48~37.5& 321.0&  758& 885&-0.70 & cj2   & S &20.0&       &1300+580\\
1310+4653&13~10~53.61&46~53~52.2& 131.0&  393& 361&-0.89 &cj2,m  & S &19.1&       &1308+471\\
1310+3233&13~10~59.45&32~33~34.9& 374.3&  688& 605&-0.50 &       & Q &19.2& 1.650 &1308+328\\
1335+4542&13~35~21.98&45~42~38.5& 251.0&  598& 468&-0.71 & cj1   & Q &18.5& 2.449 &1333+459\\
1335+5844&13~35~25.94&58~44~00.8& 292.7&  820& 766&-0.84 & cj1   &   &    &       &1333+589\\
\hline
\hline 
\end{tabular}
\end{center}
\end{table*}
}

\setcounter{table}{0}

\begin{table*}
\begin{center}
\begin{tabular}{cccrrccccclc}
\multicolumn{9}{c}{{\bf Table 1} {\it (continued)} \hfill}\\
\hline 
\hline 
J2000 &R.A.(J2000)&Dec.(J2000)&$_{NVSS}$&$_{87GB}$&$_{JVAS}$&&Other&ID&&~~~z&B1950\\ 
Name  &h~m~s  &d~m~s   & mJy  & mJy & mJy &$\alpha_{1.4}^{4.9}$&sampl.&&&&  Name  \\
\hline 
1407+2827&14~07~00.43&28~27~14.5& 817.1& 2421&1939&-0.88 &k,st,K &BG &16.0& 0.0769&1404+286\\
1410+0731&14~10~35.09&07~31~21.1& 192.3&  362& 332&-0.52 &       &   &    &       &1408+077\\
1412+1334&14~12~36.38&13~34~38.5& 196.6&  399& 242&-0.58 &       &   &    &       &1410+138\\
1419+5423&14~19~46.50&54~23~15.1& 733.9& 1707&2187&-0.69 & cj1   &BL &15.7& 0.151 &1418+546\\
1424+2256&14~24~38.13&22~56~00.6& 268.4&  503&    &-0.51 &       & Q &16.5& 3.626 &1422+231\\
1430+1043&14~30~09.78&10~43~27.1& 290.0& 1236& 822&-1.18 &  K    & Q &18.5& 1.710 &1427+109\\
1457+0749&14~57~38.09&07~49~54.0& 234.7&  618& 432&-0.79 &       &   &    &       &1455+080\\
1458+3720&14~58~44.77&37~20~22.0& 215.1&  591& 370&-0.82 &cj2,m  & G &18.2& 0.333 &1456+375\\
1505+0326&15~05~06.46&03~26~30.3& 395.4&  991& 876&-0.75 &  K    & Q &18.7& 0.411 &1502+036\\
1511+0518&15~11~41.18&05~18~09.3&  60.6&  501& 502&-1.72 &       & g &16.2& $a$   &1509+054\\
1526+6650&15~26~42.88&66~50~55.0&  88.3&  417& 312&-1.26 &cj2,m  & Q &17.2& 3.02  &1526+670\\
1551+5806&15~51~58.18&58~06~44.7& 190.5&  367& 305&-0.53 &cj2,m  ,sn & Q &15.9?& 1.324 &1550+582\\
1555+1111&15~55~43.09&11~11~24.5& 312.4&  636& 515&-0.58 &       &BL &15. & 0.360 &1553+113\\
1603+1105&16~03~41.93&11~05~49.0& 195.5&  831& 357&-1.18 &       &dss&    &       &1601+112\\
1616+0459&16~16~37.43&04~59~33.7& 352.0&  918& 693&-0.78 &  K    & Q &19.5& 3.197 &1614+051\\
1623+6624&16~23~04.44&66~24~01.0& 156.0&  520& 287&-0.98 & m,sn  & G &15.1& 0.203 &1622+665\\
%1639+1632&16~39~42.31&16~32~16.1&  12.0&  357& 166&-2.76 &       &   &    &       &1637+166\\
1645+6330&16~45~58.56&63~30~11.0& 218.2&  444& 214&-0.58 & cj2   & Q &19.4& 2.379 &1645+635\\
1716+6836&17~16~13.93&68~36~38.2& 489.3&  988& 829&-0.57 & cj2   & Q &18.5& 0.777 &1716+686\\
1717+1917&17~17~01.19&19~17~40.7& 124.6&  346& 124&-0.83 &       &   &    &       &1714+193\\
1719+0658&17~19~10.90&06~58~15.5& 117.2&  425& 107&-1.05 &       &   &    &       &1716+070\\
1722+6106&17~22~40.06&61~06~00.0& 154.8&  321& 195&-0.59 & m     & s &19.9&       &1722+611\\
1728+1215&17~28~07.03&12~15~39.2& 346.4&  958& 391&-0.83 &  K    & Q &20.0&       &1725+123\\
1735+5049&17~35~49.04&50~49~11.5& 432.0&  798& 838&-0.50 & cj1   &G? &23.1&       &1734+508\\
1740+2211&17~40~05.82&22~11~00.3& 344.7&  676& 490&-0.55 &       &g? &16.9&       &1737+222\\
1747+4658&17~47~26.68&46~58~51.0& 305.1&  634& 871&-0.60 & cj2   & S &21.3&       &1746+470\\
1751+0939&17~51~32.84&09~39~01.1& 623.1& 2283&2058&-1.06 &  k,K  & BL&16.8& 0.322 &1749+096\\
1800+3848&18~00~24.72&38~48~31.1& 326.9&  722&1177&-0.65 &cj1,k,K& Q &18  & 2.092 &1758+388\\
1811+1704&18~11~43.18&17~04~56.7& 132.5&  314& 184&-0.70 &       &   &    &       &1807+170\\
1840+3900&18~40~57.13&39~00~46.0& 143.2&  476& 221&-0.98 &cj2,m  & Q &19.5& 3.095 &1839+389\\
1849+6705&18~49~15.89&67~05~40.9& 517.9&  992& 456&-0.53 & cj2   & Q &20  & 0.657 &1849+670\\
1850+2825&18~50~27.54&28~25~12.8& 230.7&  999&1467&-1.19 &  K    & Q &17  & 2.560 &1848+283\\
1855+3742&18~55~27.65&37~42~56.0& 176.1&  341& 222&-0.54 &       &   &    &       &1853+376\\
2021+0515&20~21~35.29&05~15~05.1& 333.5&  684& 454&-0.58 &       &   &    &       &2019+050\\
2024+1718&20~24~56.47&17~18~11.3& 279.5&  586& 568&-0.60 &       & S &17.5& 1.050?&2022+171\\
2043+1255&20~43~10.18&12~55~14.0& 221.2&  429& 227&-0.54 &       &   &    &       &2040+127\\
2101+0341&21~01~38.84&03~41~32.2& 590.9& 1307& 760&-0.65 &  K    & Q &18  & 1.013 &2059+034\\
2114+2832&21~14~58.34&28~32~57.0& 396.7&  773& 486&-0.54 &       & s?&19.3&       &2112+283\\
2123+0535&21~23~44.52&05~35~22.5& 794.0& 2523&1446&-0.94 &  K    & Q &17.5& 1.878 &2121+053\\
2136+0041&21~36~38.56&00~41~54.5&3473.0&10467&7202&-0.90 &k,st,K & Q &16.8& 1.932 &2134+004\\
2203+1007&22~03~30.95&10~07~42.9& 114.6&  316& 235&-0.83 &       &   &    &       &2201+098\\
2207+1652&22~07~52.79&16~52~15.6& 208.9&  384& 280&-0.50 &       & s?&19.7&       &2205+166\\
2212+2355&22~12~06.01&23~55~40.7& 557.0& 1212& 723&-0.63 &  K    & S &19.0&       &2209+236\\
2219+1806&22~19~14.05&18~06~35.4& 159.2&  318& 357&-0.56 &       & s?&19.0&       &2216+178\\
2219+2613&22~19~49.77&26~13~27.7& 209.4&  799& 421&-1.09 &       &BG &17.0& 0.085 &2217+259\\
2230+6946&22~30~36.63&69~46~27.5& 509.1& 1365& 754&-0.80 & cj1   &BL?&19.5&       &2229+695\\
2241+4120&22~41~07.15&41~20~12.3& 336.1&  677& 826&-0.57 & cj2   & S &17.9&       &2238+410\\
2257+0243&22~57~17.54&02~43~17.5& 209.4&  426& 273&-0.58 &  K    & Q &18  & 2.081 &2254+024\\
2308+0946&23~08~44.18&09~46~26.2& 141.6&  383& 120&-0.81 &       &   &    &       &2306+095\\
2320+0513&23~20~44.83&05~13~50.5& 541.9& 1180& 387&-0.63 &  K    & Q &19  & 0.622 &2318+049\\
2321+3204&23~21~54.84&32~04~05.6& 232.9&  479& 179&-0.59 &  K    &BG &17.0&       &2319+347\\
2330+3348&23~30~13.72&33~48~36.2& 199.2&  497& 472&-0.74 &  K    & Q &18.5& 1.809 &2327+335\\
\hline
\hline 
\end{tabular}
\end{center}
\end{table*}

\clearpage

\setcounter{table}{1}

\begin{table}
\caption{VLA observations and Configurations. The total observing time
(column 3) is inclusive of the scans on the ``faint'' HFP candidates. }
\begin{center}
\begin{tabular}{cccc}
%\multicolumn{4}{c}{{\bf Table 2 }}\\
%\multicolumn{4}{l}{{\small VLA observations and Configurations }} \\
%\multicolumn{4}{c}{}\\
\hline 
\hline 
Date        & Conf. & Obs. & code \\ 
            &       & Time & \\
\hline
21~Sep~1998 &   B   & 120  &  a\\
07~Nov~1998 &  BnC  & 150  &  b\\
14~Nov~1998 &  BnC  & 150  &  c\\
19~Dec~1998 &   C   & 240  &  d\\
14~Jun~1999 &  AnD  & 420  &  e\\
21~Jun~1999 &  AnD  & 180  &  f\\
25~Jun~1999 &   A   & 120  &  g\\
25~Sep~1999 &   A   & 240  &  h\\
15~Oct~1999 &  BnA  & 240  &  i\\
25~Feb~2000 &  BnC  & 240  &  j\\
\hline 
\end{tabular}
\end{center}
\end{table}

\clearpage

\begin{table*}
\caption{Multifrequencies VLA flux densities of genuine HFP
sources.} 
\begin{center}
\begin{tabular}{ccrrrrrrrrrr}
\hline 
\hline 
J2000&Obs.&$S_{1.4}$&$S_{1.7}$&$S_{4.5}$&$S_{5.0}$&$S_{8.1}$&$S_{8.5}$&
$S_{15.0}$&$S_{22.5}$ &$S_{peak}$& $\nu_{peak}$ \\
Name     &code& mJy& mJy  & mJy & mJy & mJy & mJy & mJy & mJy & Jy & GHz \\
\hline 				
0003+2129 & a & 90 &  117 &  256&  265&  262&  257&  159&   95 & 0.27& 6.2\\
0005+0524 & e &137 &  172 &  233&  229&  189&  181&  123&  102 & 0.24& 3.4\\
0037+0808 & e & 84 &  121 &  288&  292&  260&  253&  178&  137 & 0.29& 4.9\\
0111+3906 & e & 401& 610  & 1353& 1324& 1019&  972&  500&  300 & 1.37& 4.2\\
0111+3906 & f & 388& 584  & 1343& 1320&  994&  954&  500&  301 & 1.36& 4.2\\ 
0116+2422NE& a &103 &  130 &  238&  243&  225&  220&  128&   87 & 0.24 & 4.9 \\
0217+0144 & e & 717& 841  & 1764& 1862& 2238& 2262& 2541& 2626 &$>$2.63&$>$22\\
0329+3510 & d & 391& 424  &  764&  770&  733&  731&  638&  567 & 0.79& 5.5\\
0357+2319 & d & 257& 269  &  535&  560&  605&  610&  628&  633 &$>$0.63&$>$22\\
0428+3259 & c & 149& 186  &  484&  506&  539&  536&  421&  287 & 0.55& 7.2\\
0428+3259 & d & 161& 193  &  468&  486&  525&  523&  416&  299 & 0.53& 7.6\\
0519+0848& d &183 &  174 &  141&  142&  179&  185&  292&  377 &$>$0.38&$>$22 \\
0519+0848& i &175 &  175 &  251&  278&  378&  387&  491&  560 &$>$0.56&$>$22 \\
0625+4440 & c & 194& 215  &  418&  442&  535&  540&  563&  549 & 0.57&14.5\\
0638+5933 & c & 260& 289  &  569&  591&  670&  672&  693&  667 & 0.70&12.9\\
0642+6758 & c & 232& 302  &  485&  474&  375&  366&  251&  181 & 0.50& 3.7\\
0646+4451 & c & 432& 530  & 1727& 1896& 2780& 2855& 3302& 3199 & 3.29&15.5\\
0650+6001 & c & 473& 598  & 1209& 1236& 1257& 1250& 1116&  954 & 1.27& 6.8\\
0655+4100 & c & 201& 244  &  303&  310&  333&  335&  313&  271 & 0.33& 7.8\\
0722+3722& c &151 &  173 &  233&  230&  204&  201&  148&  107 & 0.23& 4.3\\
0722+3722& f &159 &  184 &  236&  230&  205&  198&  138&   96 & 0.24& 4.1\\
0927+3902 & c &2859&3595  &10653&11308&12992&13047&12238&10948 &13.07& 8.5\\
0927+3902 & i &2772&3450  &10436&10953&11990&11859&10060& 8660 &12.15& 6.9\\
1016+0513 & d & 194& 182  &  184&  196&  306&  317&  511&  573 &$>$0.57&$>$22\\
1016+0513 & i & 294& 329  &  494&  510&  525&  522&  449&  379 & 0.52& 7.1\\
1045+0624 & d & 175& 243  &  347&  339&  300&  297&  245&  179 & 0.36& 3.7\\
1045+0624 & i & 174& 226  &  337&  331&  289&  284&  238&  196 & 0.34& 3.7\\
1148+5254 & b &  87& 100  &  354&  393&  510&  512&  501&  458 & 0.51& 8.7\\
1335+4542 & b & 252& 355  &  743&  735&  582&  564&  357&  248 & 0.75& 4.2\\
1335+5844 & e & 282& 391  &  731&  723&  680&  674&  524&  407 & 0.73& 4.9\\
1407+2827 & a & 809&1105  & 2403& 2398& 1988& 1940& 1292&  719 & 2.40& 4.9\\
1407+2827 & b & 798&1096  & 2379& 2362& 2008& 1966& 1322&  800 & 2.36& 4.9\\
1407+2827 & h & 777&1095  & 2442& 2426& 2019& 1971& 1292&  709 & 2.43& 4.9\\
1412+1334 & d & 181& 225  &  347&  330&  288&  282&  205&  147 & 0.34& 4.2\\
1424+2256 & d & 318& 393  &  620&  607&  467&  448&  270&  161 & 0.62& 4.0\\
1430+1043 & h & 288& 425  &  910&  910&  830&  816&  706&  566 & 0.91& 4.9\\
1457+0749& h &171 &  190 &  244&  241&  227&  226&  207&  177 & 0.24& 4.7\\
1505+0326 & h & 455& 547  &  921&  929&  901&  895&  830&  738 & 0.93& 6.2\\
1511+0518 & a &  66& 100  &  497&  536&  731&  738&  737&  611 & 0.77&11.0\\
1526+6650 & a & 117& 138  &  406&  411&  367&  355&  202&  109 & 0.42& 5.8\\
1603+1105& h &119 &  148 &  266&  270&  265&  264&  255&  227 & 0.27& 6.8\\
1616+0459 & h & 359& 484  &  908&  892&  698&  674&  430&  272 & 0.91& 4.1\\
1623+6624& a &148 &  178 &  289&  298&  274&  272&  229&  198 & 0.29& 5.1\\
1645+6330 & a & 221& 272  &  496&  513&  587&  588&  635&  636 &$>$0.64&$>$22\\
1717+1917& a & 87 &   98 &  192&  204&  230&  229&  227&  215 & 0.23&11.5\\
1735+5049 & a & 447& 515  &  948&  968&  935&  924&  740&  624 & 0.99& 5.9\\
1751+0939 & h &1043&1350  & 3252& 3362& 3684& 3690& 3691& 3367 & 3.70& 8.5 \\
1751+0939 & e &1399&1673  & 3525& 3688& 4201& 4194& 4178& 4031 & 4.24&10.7\\
1800+3848 & e & 270& 328  &  733&  791& 1101& 1125& 1364& 1363 & 1.40&17.8\\
1811+1704 & h & 286& 381  &  678&  691&  763&  765&  837&  795 & 0.82&14.7\\
1840+3900& e &152 &  168 &  204&  203&  195&  191&  183&  167 & 0.20& 4.5\\
1850+2825 & e & 210& 284  & 1135& 1246& 1561& 1550& 1301& 1002 & 1.56& 8.3\\
1855+3742 & e & 191& 186  &  384&  364&  234&  221&  136&  106 & 0.39& 4.5\\
\hline
\hline 
\end{tabular}
\end{center}
\end{table*}

\setcounter{table}{2}

\begin{table*}
\caption{Multifrequencies VLA flux densities of genuine HFP
sources. {\it (Continued)}} 
\begin{center}
\begin{tabular}{ccrrrrrrrrrr}
\hline 
\hline 
J2000&Obs.&$S_{1.4}$&$S_{1.7}$&$S_{4.5}$&$S_{5.0}$&$S_{8.1}$&$S_{8.5}$&
$S_{15.0}$&$S_{22.5}$ &$S_{peak}$& $\nu_{peak}$ \\
Name     &code& mJy& mJy  & mJy & mJy & mJy & mJy & mJy & mJy & Jy & GHz \\
\hline 				
2021+0515 & f & 331& 383  &  492&  477&  387&  372&  251&  158 & 0.49& 3.7\\
2024+1718 & e & 268& 304  &  544&  572&  740&  745&  838&  776 & 0.84&14.5\\
2101+0341 & e & 481& 534  &  854&  920& 1283& 1298& 1545& 1583 &$>$1.58&$>$22\\
2114+2832 & e & 371& 396  &  691&  722&  797&  792&  749&  685 & 0.79& 9.8\\
2123+0535 & e & 861&1147  & 1876& 1896& 2121& 2135& 2515& 2559 &$>$2.56&$>$22\\
2136+0041 & a &3682&4976  & 9514& 9537& 8532& 8370& 6475& 5412 & 9.57& 4.5\\
2136+0041 & d &3478&4811  & 9685& 9702& 9129& 8940& 7443& 6169 & 9.71& 5.0\\
2203+1007 & e & 103& 155  &  321&  319&  258&  247&  130&   89 & 0.33& 4.2\\
2207+1652 & e & 193& 256  &  538&  551&  554&  547&  477&  417 & 0.56& 6.3\\
2212+2355 & e & 665& 739  & 1141& 1182& 1334& 1337& 1374& 1280 & 1.38&12.6\\
2257+0243& e &181 &  182 &  251&  274&  412&  425&  576&  575 & 0.58&19.5\\
2320+0513 & e & 665& 839  & 1197& 1196& 1088& 1069&  885&  755 & 1.21& 4.1\\
2330+3348 & e & 388& 439  &  551&  558&  559&  520&  479&  388 & 0.59& 5.6\\
\hline
\hline 
\end{tabular}
\end{center}
\end{table*}

\begin{table*}
\caption{Non HFP sources.}
\begin{center}
\begin{tabular}{ccrrrrrrrrrr}
\hline 
\hline 
J2000&Obs.&$S_{1.4}$&$S_{1.7}$&$S_{4.5}$&$S_{5.0}$&$S_{8.1}$&$S_{8.5}$&
$S_{15.0}$&$S_{22.5}$ \\
Name     &code& mJy& mJy  & mJy & mJy & mJy & mJy & mJy & mJy \\
\hline 				
0037+1109& a & 223&  232 &  228&  224&  203&  198&  164&  146 \\
0039+1411& a & 262&  282 &  290&  295&  279&  273&  239&  223 \\
0107+2611& a & 195&  184 &  121&  118&   99&   98&   74&   53 \\
0116+2422SW& a & 142&  133 &  158&  158&  161&  165&  172&   183\\
0132+4325& e & 221&  240 &  333&  348&  401&  402&  388&  332 \\
0205+1444& e & 156&  170 &  175&  174&  161&  159&  123&  113 \\
0254+3931& f & 208&  212 &  266&  275&  335&  334&  348&  355 \\
0310+3814& d & 556&  545 &  588&  584&  543&  542&  525&  512 \\
0312+0133& e & 362&  377 &  265&  250&  191&  189&  162&  159 \\
0313+0228& e & 183&  198 &  219&  226&  260&  264&  281&  272 \\
0424+0036& e & 442&  484 &  638&  667&  783&  794&  842&  873 \\
0509+0541& d & 608&  609 &  607&  616&  656&  663&  645&  632 \\
0530+1331& c &2145& 2281 & 2946& 3056& 3476& 3475& 3294& 3120 \\
0530+1331& i &2102& 2486 & 2905& 2876& 2969& 3031& 3944& 4428 \\
0559+5804& f & 306&  313 &  304&  304&  312&  313&  312&  283 \\
0733+0456& d & 194&  208 &  265&  268&  291&  295&  323&  322 \\
0811+0146& d & 660&  722 &  863&  875&  908&  913&  888&  828 \\
0831+0429& d & 879&  850 &  822&  823&  836&  838&  841&  839 \\
0854+0720& d &  97&   97 &  113&  113&  118&  119&  118&  114 \\
0958+6533& c & 392&  390 &  335&  333&  326&  325&  334&  322 \\
1018+0530& d & 291&  244 &  190&  183&  155&  152&  144&  141 \\
1033+0711& d & 273&  298 &  384&  390&  408&  406&  378&  322 \\
1048+7143& b &1479& 1544 & 1653& 1656& 1573& 1564& 1474& 1419 \\
1056+7011& b & 312&  306 &  353&  352&  309&  302&  252&  263 \\
1146+3958& b & 524&  576 &  939&  958&  982&  980&  955&  971 \\
1146+3958& i & 589&  644 &  760&  759&  735&  735&  803&  912 \\
1209+4119& b & 155&  162 &  182&  184&  172&  171&  157&  154 \\
1228+3706& b & 298&  304 &  408&  418&  434&  430&  364&  307 \\
1302+5748& e & 245&  270 &  399&  412&  471&  464&  494&  454 \\
1310+4653& b & 144&  152 &  195&  201&  199&  196&  165&  154 \\
1310+3233& d & 298&  298 &  350&  354&  383&  385&  417&  422 \\
1410+0731& d & 160&  155 &  145&  147&  154&  155&  161&  152 \\
1419+5423& j & 604&  610 &  752&  756&  761&  763&  770&  760 \\
1458+3720& g & 155&  161 &  185&  184&  181&  180&  180&  125 \\
1551+5806& h & 202&  222 &  262&  256&  244&  242&  275&  304 \\
1555+1111& a & 271&  280 &  245&  249&  241&  238&  228&  216 \\
%1639+1632& a & 118&  125 &  112&  114&  113&  112&  106&  100 \\
1716+6836& a & 448&  449 &  561&  575&  631&  634&  678&  692 \\
1719+0658& h & 245&  260 &  393&  393&  383&  379&  379&  359 \\
1722+6106& a & 182&  179 &  142&  142&  151&  154&  189&  181 \\
1728+1215& h & 240&  260 &  387&  397&  454&  459&  480&  453 \\
1740+2211& a & 304&  303 &  319&  322&  282&  275&  211&  180 \\
1747+4658& e & 340&  350 &  387&  390&  393&  389&  365&  316 \\
1849+6705& f & 481&  471 &  459&  470&  563&  567&  707&  555 \\
2043+1255& e & 184&  199 &  233&  236&  245&  241&  221&  193 \\
2219+1806& e & 133&  141 &  172&  174&  206&  208&  214&  206 \\
2219+2613& e & 157&  167 &  199&  199&  175&  168&  136&  148 \\
2230+6964& e & 397&  381 &  350&  355&  385&  392&  444&  403 \\ 
2241+4120& e & 426&  420 &  313&  309&  280&  279&  236&  238 \\
2308+0946& a & 209&  213 &  190&  192&  218&  220&  239&  230 \\
2321+3204& a & 375&  397 &  413&  418&  407&  408&  372&  368 \\
\hline
\hline 
\end{tabular}
\end{center}
\end{table*}

\begin{figure*}
\includegraphics{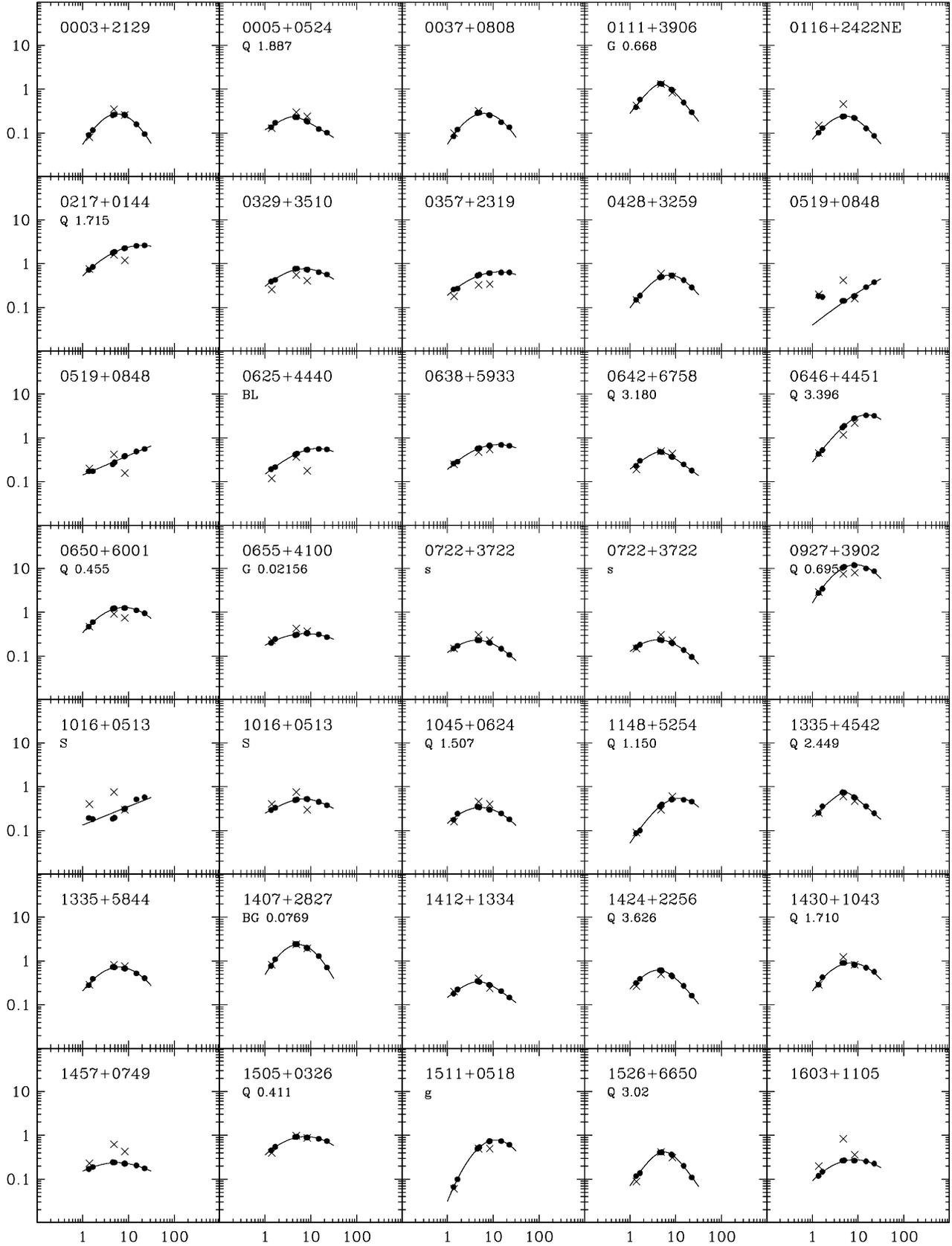}
\vspace{22.3cm}
\caption{Spectra of genuine HFP sources: filled circles represent the
simultaneous multifrequency VLA data, while crosses are used for the flux
densities from the catalogues mentioned in the text. The solid line
shows the fitting curve. Log($\nu$) (GHz) and Log($F_\nu$) (Jy) are
the $x$ and $y$ axis respectively. }
\label{fig_HFPok}
\end{figure*}

\clearpage

\setcounter{figure}{0}

\begin{figure*}
\includegraphics{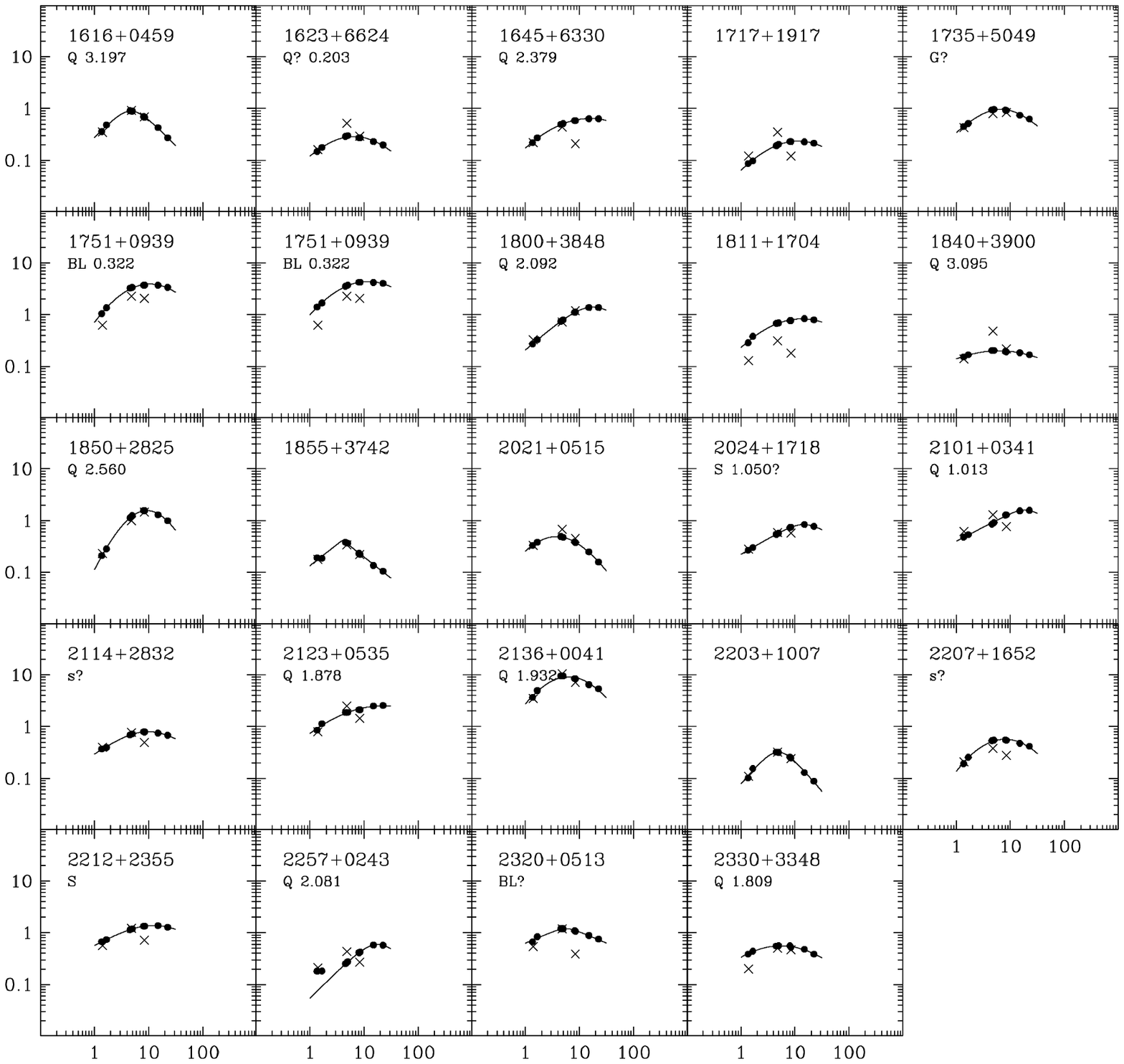}
\vspace{16.5cm}
\caption{Spectra of genuine HFP sources $(Continued)$.}
\label{fig_HFPok2}
\end{figure*}

\begin{figure*}
\includegraphics{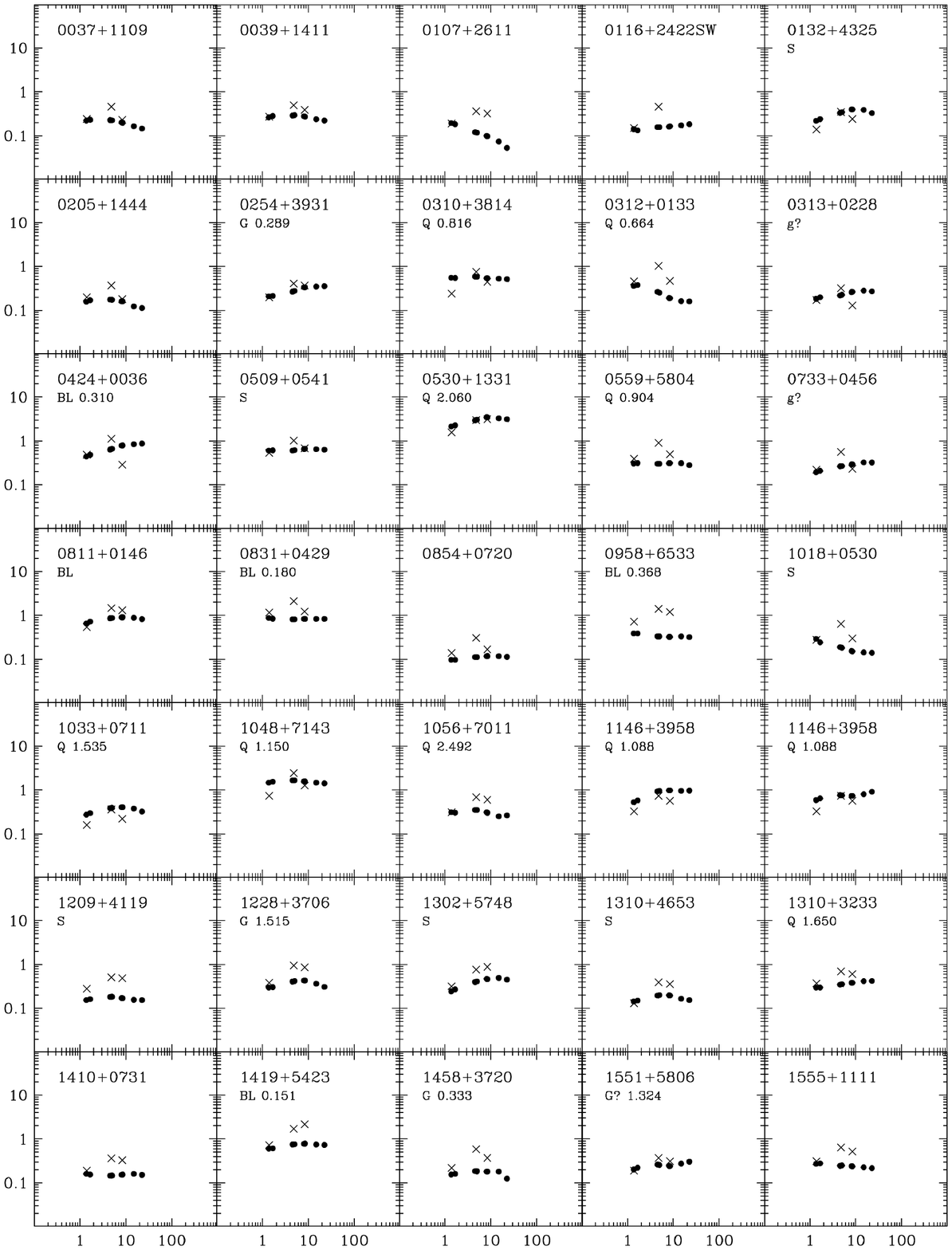}
\vspace{22.3cm}
\caption{Spectra of non HFP sources: as for Fig. 1, filled circles
represent the simultaneous multifrequency VLA data, while crosses are
used for the flux densities from the catalogues mentioned in the
text. Log($\nu$) (GHz) and Log($F_\nu$) (Jy) are the $x$ and $y$ axis
respectively. } 
\label{fig_nonHFP1}
\end{figure*}

\setcounter{figure}{1}

\begin{figure*}
\includegraphics{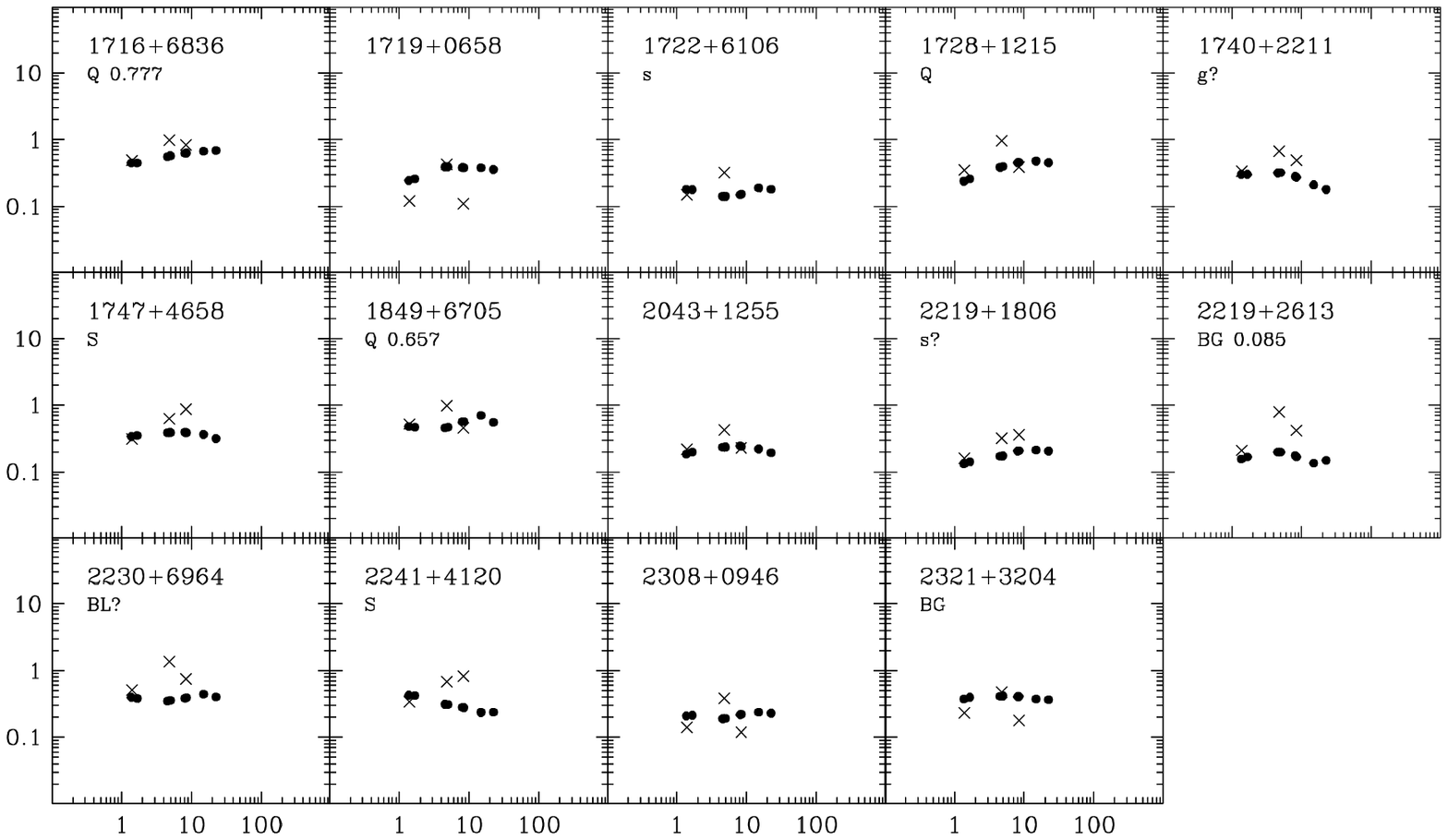}
\vspace{9.6cm}
\caption{Spectra of non HFP sources $(Continued)$.}
\label{fig_nonHFP2}
\end{figure*}

\clearpage

\begin{figure*}
\includegraphics{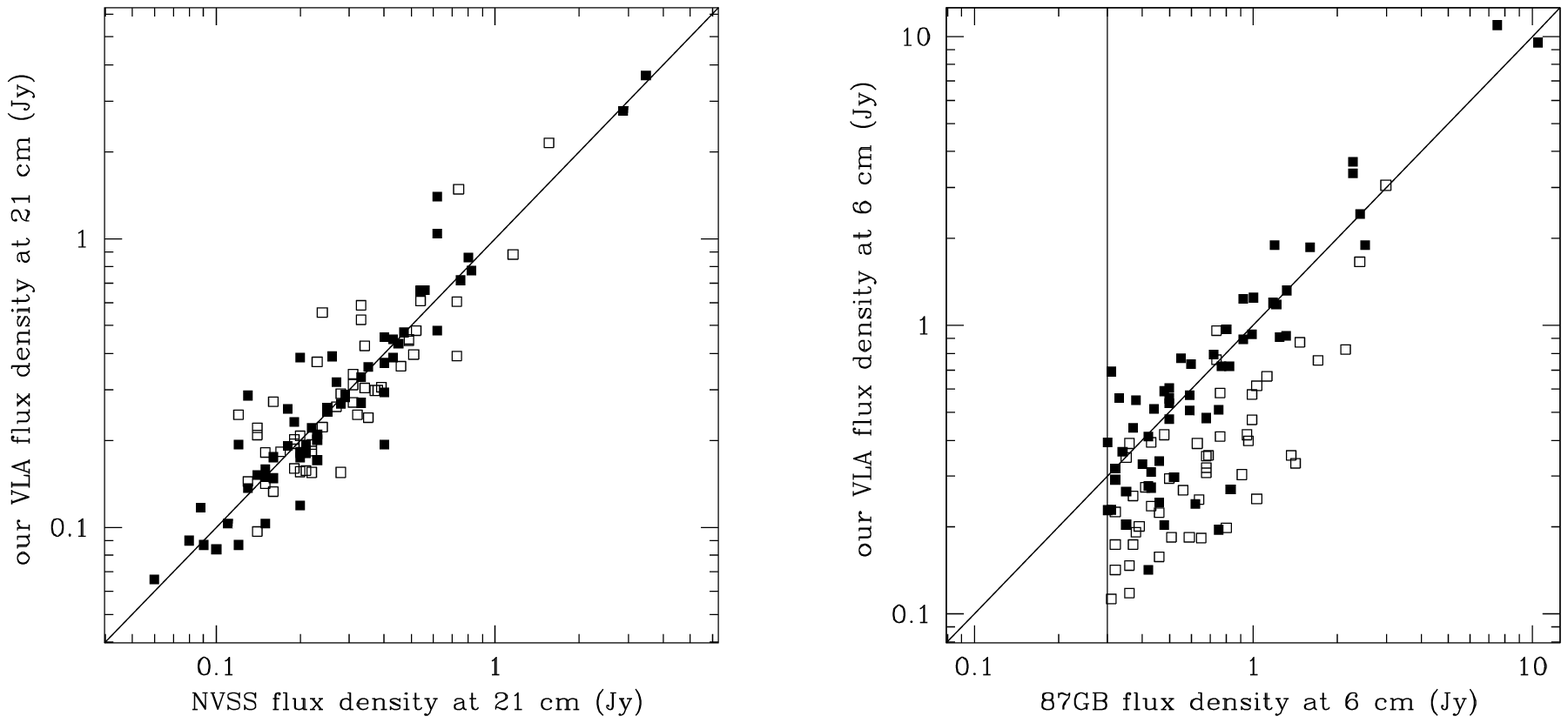}
\vspace{9.6cm}
\caption{Flux densities from the NVSS ({\it left}) and 87GB ({\it
right}) catalogues are compared to ours at the same frequency. Filled
squares represent genuine HFP sources (both ``bright'' and ``weak'',
while open squares are for non HFPs.} 
\label{fig_comparison}
\end{figure*}

\clearpage

\setcounter{figure}{2}

\begin{figure}
\includegraphics{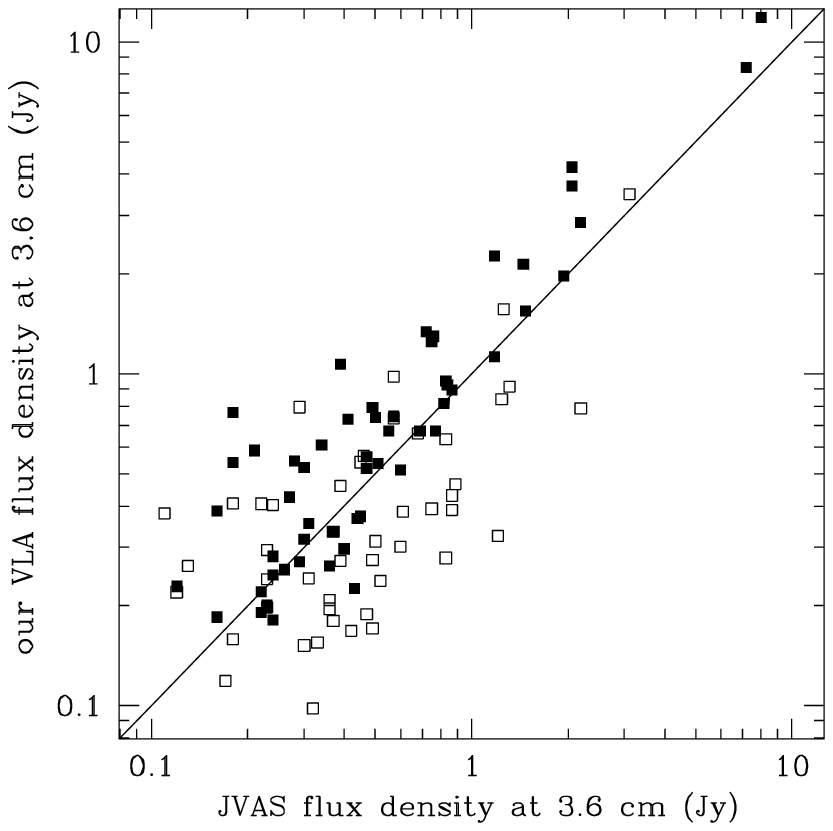}
\vspace{9.6cm}
\caption{$(Continued)$ Comparison between our flux density measures at
8.46 GHz  with the Flux densities from the JVAS catalogue. Filled
squares represent genuine HFP sources (both ``bright'' and ``weak''),
while open squares are for non HFPs.}
\label{fig_comparison}
\end{figure}

\listofobjects

\end{document}